\documentclass[12pt]{article}%
\usepackage{amsfonts}
\usepackage{amssymb}
\usepackage{amsmath}
\usepackage{geometry}
\usepackage{graphicx}
\usepackage{lscape}%
\providecommand{\U}[1]{\protect\rule{.1in}{.1in}}

\begin{document}

\title{ }

\begin{center}
\bigskip\textit{Original Manuscript}

\bigskip{}

{\LARGE The Nudge Average Treatment Effect }

\bigskip{\Large Eric Tchetgen Tchetgen }

Department of Statistics and Data Science

Department of Biostatistics,\ Epidemiology and Informatics

University of Pennsylvania\bigskip\bigskip

\textbf{Abstract}
\end{center}

\noindent The instrumental variable method is a prominent approach to obtain
under certain conditions, valid inferences about a treatment causal effect
even when unmeasured confounding might be present. In the highly celebrated
paper, Imbens and Angrist (1994) established that a valid instrument
nonparametrically identifies the average causal effect among compliers, also
known as the local average treatment effect under a certain monotonicity
assumption which rules out the existence of so-called defiers. An often-cited
attractive property of monotonicity is that it facilitates a causal
interpretation of the instrumental variable estimand without restricting the
degree of heterogeneity of the treatment causal effect. In this paper, we
introduce an alternative equally straightforward and interpretable condition
for identification, which accommodates both the presence of defiers and
heterogenous treatment effects. Mainly, we show that under our new conditions,
the instrumental variable estimand recovers the average causal effect for the
subgroup of units for whom the treatment is manipulable by the instrument, a
subgroup which may consist of both defiers and compliers, therefore recovering
an effect estimand we aptly call the \textit{Nudge Average Treatment Effect}.

\newpage

\section{Introduction}

\noindent Confounding is a major concern when drawing causal inference from
observational data. For this reason, when estimating the effect of a treatment
of interest, one must endeavor to adjust, to the extent that they are
observed, for as many potential confounders as practically feasible.
Nonetheless, despite such efforts, unmeasured confounding can seldom be ruled
out with certainty in non-experimental studies. The instrumental variable (IV)
method offers a principled, easily interpretable approach to recover under
certain conditions, valid inferences about a treatment causal effect even when
unmeasured confounding persists. The core assumptions of an instrumental
variable is that in addition to being relevant for the treatment, it does not
have a direct effect on the outcome, other than through the treatment, and
that it must be independent of any unmeasured confounder. \ In their highly
celebrated paper, Imbens and Angrist (1994) established the nonparametric
identification of the so-called Local Average Treatment Effect (LATE),
corresponding to the average treatment effect for a subgroup of units known as
compliers, under a monotonicity condition relating the instrument and
treatment variables which rules out the presence of defiers; also see Permutt
and Hebel (1989) and Baker and Lindeman (1994) for contemporaneous related
developments. An appealing and often-cited feature of the LATE\ framework is
that it provides an easily interpretable condition for obtaining a model-free
causal interpretation of the instrumental variable estimand without the need
to restrict the degree of treatment effect heterogeneity. \ 

In this paper, we introduce an alternative equally straightforward and
interpretable instrumental variable identification condition which readily
accommodates both the presence of defiers and unrestricted treatment
heterogeneity. Throughout, we define as \textit{nudge-able}, a unit whose
treatment selection is prone to change in response to a nudge induced by the
instrument, irrespective of whether the specific nudge induces what is
commonly characterized in the IV literature (Imbens and Angrist, 1994) as
compliant or defiant treatment uptake behavior under potential instrumental
variable assignments. \ Importantly, "nudge-ability" is relative to a specific
instrument for a specific treatment under consideration, and not necessarily
an immutable personal characteristic, in that a person's treatment uptake may
be manipulable by one instrument but may not by another.

The new condition essentially states that within the subgroup of nudge-able
units, any heterogeneity of the treatment effect induced by hidden
counfounders must be uncorrelated with corresponding heterogeneity in the
share of compliers. An important and easily interpretable sufficient condition
for this property is that, although a priori unrestricted, the share of
compliers within the subgroup of nudge-able units is balanced across strata of
the unmeasured confounders. Notably, our novel condition is implied by
monotonicity which we recover as a special case where the share of compliers
entails the entire nudge-able population across strata of the unmeasured
confounders. Importantly, the converse is generally not true. Therefore one
may view our proposed condition as substantially less restrictive than
monotonicity, while preserving a formal and straightforward interpretation of
the IV estimand as an average causal effect for the subgroup of units for whom
the treatment can be manipulated by the instrument, i.e. the nudge-able
subgroup which may be arbitrarily composed of both defiers and compliers,
motivating the label of the identified treatment effect as the\textit{ Nudge
Average Treatment Effect }(NATE).

\ To facilitate interpretability, and to further provide intuition for the
proposed identification strategy, it will be useful to consider a
generalization of a certain treatment selection model first proposed by
Heckman (1976), known as a latent index model (LIM). \ Under Heckman's LIM,
one models a unit's treatment selection with a latent index crossing a random
threshold, where the latent index represents a form of expected net utility of
selecting into treatment. Interestingly, Vytlacil (2002) established a
fundamental equivalence between Heckman's LIM and the LATE assumptions of
Imbens and Angrist (1994), therefore offering an alternative formulation of
the LATE identifying model. Later, we adopt a generalization of LIM, we refer
to as Generalized LIM (GLIM)\ to relate IV models for the various prominent
causal effects in the literature, mainly the Average Treatment effect (ATE),
the Average Treatment Effect on the Treated (ATT) and NATE. Thus, we establish
that GLIM encompasses: (i) Heckman's LIM (and therefore the LATE model of
Imbens and Angrist (1994)) as a special case; (ii) a class of selection models
nested within the additive treatment selection IV\ model (AIV) of Wang and
Tchetgen Tchetgen (2018), which identifies the population ATE; (iii)\ a class
of selection models nested within the Multiplicative treatment selection
IV\ model (MIV) of Liu et al (2025), which identifies the average treatment
effect for the treated (ATT); and (iv) a class of treatment selection models
nested within our proposed IV model for identifying the NATE. Therefore, the
GLIM\ framework provides a useful platform to reason, articulate and
communicate about similarities and differences of three prominent IV\ models
(i)-(iii) in the literature, particularly in relation to our proposed NATE
model (iv).

\section{Notation and Definitions}

Throughout, we let $Z$ denote a binary instrumental variable, and $\left\{
A^{z=0},A^{z=1}\right\}  $ denote binary counterfactual treatment variables
under an external intervention which sets the instrument to $z=0,1$,
respectively. Throughout, we assume that the intervention on the instrument
giving rise to these counterfactual treatment variables are well-defined and
could potentially be performed in the context of a randomized experiment, at
least in principle if not in actually. \ Let $\mathcal{C}$ denote a unit's
compliance type defined in terms of the joint counterfactual treatments
$\left\{  A^{z=0},A^{z=1}\right\}  $ as followed
\[
\mathcal{C}=\left\{
\begin{array}
[c]{c}%
\mathrm{nt}\text{ if }A^{z=0}=0,A^{z=1}=0;\\
\mathrm{at}\text{ if }A^{z=0}=1,A^{z=1}=1;\\
\mathrm{de}\text{ if }A^{z=0}=1,A^{z=1}=0;\\
\mathrm{co}\text{ if }A^{z=0}=0,A^{z=1}=1;
\end{array}
\right.
\]
where units of compliance type $\left\{  \mathcal{C}=\mathrm{nt}\right\}  $
are so-called never-takers, $\left\{  \mathcal{C}=\mathrm{at}\right\}  $ are
always takers, $\left\{  \mathcal{C}=\mathrm{de}\right\}  $ are defiers, and
$\left\{  \mathcal{C}=\mathrm{co}\right\}  $ are compliers using terminology
due to Imbens and Angrist (1994) and Angrist, Imbens and Rubin (1996). In the
following, let $\mathrm{I}\left\{  \mathcal{E}\right\}  $ denote the indicator
for the event $\mathcal{E}$. \ The observed treatment is obtained by the
consistency assumption as followed:
\begin{align*}
A  &  =Z\left(  \mathrm{I}\left\{  \mathcal{C}=\mathrm{at}\right\}
+\mathrm{I}\left\{  \mathcal{C}=\mathrm{co}\right\}  \right)  +\left(
1-Z\right)  \left\{  \mathrm{I}\left\{  \mathcal{C}=\mathrm{at}\right\}
+\mathrm{I}\left\{  \mathcal{C}=\mathrm{de}\right\}  \right\} \\
&  =ZA^{z=1}+(1-Z)A^{z=0}\\
&  =A^{Z}.
\end{align*}
\newline Likewise, let $\left\{  Y^{a=1},Y^{a=0}\right\}  $ denote the
potential outcomes for treated and control states, respectively, and the
observed outcome is obtained by the corresponding consistency assumption
$Y=AY^{a=1}+(1-A)Y^{a=0}.$ Throughout, let $L$ stand for observed confounders
which may be associated with the instrument, the treatment, the outcome and
any unmeasured confounder.

\section{The LATE Approach}

We briefly review identification of the LATE under the following conditions of
Imbens and Angrist (1994), we now state allowing for potential confounding\ of
the effects of the instrument by $L$:%
\begin{align}
&  Z\amalg\left(  Y^{a=1},Y^{a=0},A^{z}\right)  |L\label{IA Independence}\\
A^{z=0}  &  \leq A^{z=1}\text{ for every unit}\label{IA Monotonicity}\\
\Pr\left(  A^{z=1}=1|L=l\right)  -\Pr\left(  A^{z=0}=1|L=l\right)   &
\neq0\text{ for every value of }l \label{IA relevance}%
\end{align}
Assumption $\left(  \ref{IA Independence}\right)  $ implies that conditional
on $L,$ (i) the causal effects of $Z$ on $A$ and $Y$ is unconfounded; and (ii)
$Z$ does not have a direct effect on $Y$ other than possibly through $A,$ also
known as the exclusion restriction. Assumption $\left(  \ref{IA Monotonicity}%
\right)  $ is a monotonicity condition which rules out the existence of
defiers, i.e%
\[
\Pr\left\{  \mathcal{C=}\mathrm{de}|l\right\}  =\Pr\left\{  A^{z=0}%
=1,A^{z=1}=0|l\right\}  =0\text{ for all }l;
\]
while condition $\left(  \ref{IA relevance}\right)  $ ensures IV\ relevance
for the treatment within levels of $L,$ that the IV is associated with the
treatment in all strata of the observed covariates. For any subset of
covariates $V$ such that $L=\left(  V^{\prime},W^{\prime}\right)  ^{\prime}$,
one may define the conditional LATE given $V=v$ as the average causal effect
for compliers with $V=v:$%
\[
\beta\left(  v\right)  =E\left(  Y^{a=1}-Y^{a=0}|\mathcal{C=}\mathrm{co}%
\text{,}V=v\right)
\]
Then, under conditions $\left(  \ref{IA Independence}\right)  $-$\left(
\ref{IA relevance}\right)  ,$ following a similar argument as Imbens and
Angrist (1994), the conditional LATE is identified by:
\[
\beta\left(  v\right)  \equiv\frac{E\left(  Y^{z=1}|V=v\right)  -E\left(
Y^{z=0}|V=v\right)  }{\Pr\left(  A^{z=1}=1|V=v\right)  -\Pr\left(
A^{z=0}=1|V=v\right)  }%
\]
where $Y^{z}\equiv Y^{A^{z}}$ and
\[
E\left(  Y^{z}|V=v\right)  =\int E\left(  Y|Z=z,L=l\right)  dF\left(
l|v\right)
\]
and
\[
E\left(  A^{z}|V=v\right)  =\int E\left(  A|Z=z,L=l\right)  dF\left(
l|v\right)
\]
are given by the standard g-formula (Robins, 1986). In the simple case where
all conditions hold without conditioning on $L,$ one obtains the standard Wald
ratio:
\[
\beta=E\left(  Y^{a=1}-Y^{a=0}|\mathcal{C=}\mathrm{co}\right)  =\frac{E\left(
Y|Z=1\right)  -E\left(  Y|Z=0\right)  }{\Pr\left(  A=1|Z=1\right)  -\Pr\left(
A=1|Z=0\right)  }.
\]

\section{The NATE\ Approach}

We next define the Nudge Average Treatment Effect as the average causal effect
for the subgroup of units whose treatment is nudge-able. In this vein, let
\begin{align*}
\mathcal{N}  &  \mathcal{=}\mathrm{I}\left\{  A^{z=1}\not =A^{z=0}\right\} \\
&  =\mathrm{I}\left\{  A^{z=1}>A^{z=0}\right\}  +\mathrm{I}\left\{
A^{z=1}<A^{z=0}\right\}
\end{align*}
denote an indicator of whether a person's treatment is nudge-able, then the
conditional NATE given $V=v$ is defined as:%
\[
\psi\left(  v\right)  \equiv E\left(  Y^{a=1}-Y^{a=0}|\mathcal{N}%
=1,V=v\right)
\]
In case monotonicity holds, the NATE naturally reduces to the LATE, i.e.
$\beta\left(  v\right)  =\psi\left(  v\right)  ,$ however, the two effects
will generally not coincide in absence of monotonicity, in which case, the
NATE may be of valuable independent interest.

Before describing our identification conditions, we briefly discuss two
examples motivating the practical relevance of the NATE, particularly in
settings where monotonicity may not be realistic. \ For instance, a violation
of the monotonicity assumption is nicely motivated in the original Example 2
of Imbens and Angrist (1994) in which two officials are screening applicants
for a social program, the causal impact of which on a welfare outcome is of
scientific interest. As they describe, we would expect that for any set of
measured characteristics, the admission rate likely differs between the two
officials. When the identity of the official cannot plausibly causally affect
the outcome of participation or nonparticipation in the program, then,
\textit{official identity} qualifies as an instrument. Supposing the admission
rate for official A was higher than for official B; then, monotonicity holds
whenever any applicant who would have been accepted by official B is accepted
by official A. Imbens and Angrist (1994) note that \textquotedblleft this is
unlikely to hold if admission is based on a number of
criteria.\textquotedblright\ In this case, monotonicity is likely violated,
and the LATE\ (i.e. the average causal effect for individuals who would only
enroll in the program if accepted by official A but not otherwise) may not be
identified. Instead, one may consider the average causal effect for
individuals who would be accepted in the program only if screened by one of
the two officials, but not the other, irrespective of their preference; this
is the average causal effect in the subset of individuals whose treatment
status can be manipulated by a randomized nudge induced by an official's
identity, the NATE. A related discussion pertaining to physician prescribing
preferences as instrumental variables for which monotonicity may be
unrealistic in comparative effectiveness studies, can be found in Swanson et
al (2015) and Boef et al (2016).

Likewise, in Mendelian randomization studies, it is likely that while a
genetic variant (say the FTO gene) may cause the phenotype defining the
exposure of interest (say obesity status in early adolescence), several other
individual characteristics (a large number of factors related to diet,
exercise, substance abuse and other lifestyle choices) are likely also strong
determinants of obesity, such that, even if the exclusion restriction holds
for the outcome of interest\ (say depression diagnosis in early adulthood),
and thus the genetic variant is a valid instrument, monotonicity would require
that a person who is obese but does not carry a minor allele of FTO, would
have necessarily been obese, had they carried the minor allele. \ This may be
unrealistic given the large number of factors that may determine a person's
obesity status in early adolescence. In this case, the NATE corresponds to the
causal effect of genetically induced obesity; that is, it is the causal effect
associated with a randomized genetic nudge of a person's obesity status.

To state our main identification result, we will let $U$ denote a set of
unmeasured confounders of the causal effect of $A$ on $Y.$ The NATE model
involves a somewhat different formulation of the instrumental variable model
than Imbens and Angrist (1994), as one that satisfies the following
conditions:
\begin{align}
&  Z\amalg\left(  U,Y^{a=1},Y^{a=0},\mathcal{C}\right)
|L;\label{independence}\\
&  \mathcal{C}\amalg\left(  Y^{a=1},Y^{a=0}\right)
|L,U;\label{Latent Unconfoundedness}\\
\Pr\left(  A^{z=1}=1|u,l\right)  -\Pr\left(  A^{z=0}=1|u,l\right)   &
\not =0\text{ for all }u,l. \label{relevance}%
\end{align}

Assumption $\left(  \ref{independence}\right)  $ implies that conditional on
$L,$ (i) the causal effects of $Z$ on $A$ and $Y$ is unconfounded$;$ (ii) $Z$
is independent of the unmeasured confounder $U;$ and (iii) $Z$ does not have a
direct effect on $Y$ other than through $A$, i.e. the exclusion restriction.
Likewise, Assumption $\left(  \ref{Latent Unconfoundedness}\right)  $ implies
that adjusting for $L$ and $U$ identifies the causal effect of $A$ on $Y,$
while throughout, adjusting for $L$ only may not be sufficient for
identification. Specifically, the condition implies the familiar latent
unconfoundedness condition,
\[
A\amalg\left(  Y^{a=1},Y^{a=0}\right)  |L,U;
\]
for example, see Wang and Tchetgen Tchetgen (2018) and Liu et al (2025). The
condition resembles a standard unconfoundedness condition had, contrary to
fact, all relevant confounders been observed. This assumption is quite mild,
as it only requires such a set of variables exists, even if unobserved. The
last assumption $\left(  \ref{relevance}\right)  $ extends IV\ relevance to
all joint levels of $L$ and $U;$ as pointed out by an observant reviewer, this
assumption rules out the existence of a stratum of $(U,L)$ in which the
proportions of compliers and defiers are equal. That is, within levels of the
covariates (both measured and unmeasured), there must be at least one more
individual who is partial to either being a complier (defier)\ than to being a
defier\ (complier). This assumption is of course implied by monotonicity but
needs to be highlighted in its absence.

To state our main identification result, we will also need to consider:%
\begin{align*}
\Delta_{y}\left(  U,L\right)   &  \equiv E\left(  Y^{a=1}-Y^{a=0}%
|\mathcal{N}=1,U,L\right) \\
\pi\left(  U,L\right)   &  \equiv\Pr\left\{  \mathcal{C=}\mathrm{co}%
|\mathcal{N}=1,U,L\right\}
\end{align*}
The function $\Delta_{y}\left(  u,l\right)  $ defines the conditional NATE for
the subgroup of units with $\left\{  U=u,L=l\right\}  ,$ while $\pi\left(
u,l\right)  $ defines the share of compliers in the subgroup of nudge-able
units with $\left\{  U=u,L=l\right\}  .$ \ Our results are based on the
following condition:%

\begin{equation}
\mathrm{COV}\left(  \Delta_{y}\left(  U,L\right)  ,\pi\left(  U,L\right)
|\mathcal{N}=1,V\right)  =0,\text{ } \label{null cov}%
\end{equation}
where $\mathrm{COV}(X,M|R)$ stands for the conditional covariance between $X$
and $M$ given $R,$ and $V$ is a specified subset of the measured covariates,
that is $L=\left(  V^{\prime},W^{\prime}\right)  ^{\prime}$. The condition
essentially states that within levels of\ $V$ among nudge-able units,
heterogeneity in the causal effect induced by variability in $(U,W)$ is
uncorrelated with corresponding heterogeneity in the share of compliers.\ For
instance, the condition would naturally be satified for $V=L,$ under
homogeneous treatment effects with respect to unmeasured confounders, i.e. if
$E\left(  Y^{a=1}-Y^{a=0}|\mathcal{N}=1,U,L\right)  =E\left(  Y^{a=1}%
-Y^{a=0}|\mathcal{N}=1,L\right)  .$ In our Mendelian randomization example,
this would correspond the additive average causal effect of adolescence
obesity status on early adulthood depression status (as measured by a standard
depression screening score such as the Patient Health Questionnaire-9) being
independent of any hidden factor conditional on measured covariates.
Alternatively, a sufficient condition for $\left(  \ref{null cov}\right)
\ $which does not restrict the degree of effect heterogeneity entails the
following\textit{ balanced complier's share (BCS)} condition, conditional on
$L$:%
\begin{equation}
\mathcal{C}\amalg U|\mathcal{N}=1,L \label{BCS}%
\end{equation}
in which case there exists a function $\widetilde{\pi}(L)$ such that
$\widetilde{\pi}(L)$ $=\pi\left(  U,L\right)  ;$ then it immediately follows
that $\left(  \ref{null cov}\right)  $ holds with $V=L$ because the share of
nudge-able units that are compliers remains constant across levels of $U,$
conditional on $L,$ and therefore by definition, is uncorrelated with any
heterogeneity of the conditional treatment effect that might be induced by
$U$. Note that while the assumption implies that the share of compliers is
balanced across strata of $U$ among nudge-able units with the same value of
$L,$ the share of compliers may vary arbitrarily across levels of $L$ without
compromising identification$.$ In the Mendelian randomization example, this
would correspond to an assumption that while hidden biological or lifestyle
confounding factors $U$ may determine whether a person's obesity status might
be genetically determined by FTO, they do not differentially determine whether
the genetic variant instrument increases or decreases such a person's average
risk for obesity. A\ specific generative model compatible with this hypothesis
is further discussed in Section 6.

Clearly, monotonicity implies $\left(  \ref{BCS}\right)  $ and therefore
condition $\left(  \ref{null cov}\right)  $, since then $\pi\left(
U,L\right)  \equiv1$ and therefore is uncorrelated with $\Delta_{y}\left(
U,L\right)  .$ \ It is notable that $\left(  \ref{null cov}\right)  $ can
generally hold even if monotonicity fails as in the case of BCS$.$ Another
interesting case where $\left(  \ref{null cov}\right)  $ holds without ruling
out latent heterogeneity in either $\pi$ or $\Delta_{y}$ with respect to $U$,
is if there exists functions $\overline{\Delta}_{y}\left(  U_{1},V\right)  $
and $\overline{\pi}\left(  V,U_{2}\right)  $ such that $\Delta_{y}\left(
U,L\right)  =\overline{\Delta}_{y}\left(  U_{1},V\right)  $ and $\pi\left(
U,L\right)  =\overline{\pi}\left(  V,U_{2}\right)  $ with $U_{1},U_{2}$
separate components of $U$ that satisfy $U_{1}\amalg U_{2}|V$. Our first main
result follows:

\noindent\textit{Result 1: Suppose that conditions} $\left(
\ref{independence}\right)  ,\left(  \ref{Latent Unconfoundedness}\right)
,\left(  \ref{relevance}\right)  $ \textit{and }$\left(  \ref{null cov}%
\right)  $\textit{ hold}, \textit{then}%

\begin{equation}
\psi\left(  v\right)  =E\left(  Y^{a=1}-Y^{a=0}|\mathcal{N}=1,v\right)
=\frac{E\left(  Y^{z=1}|v\right)  -E\left(  Y^{z=0}|v\right)  }{\Pr\left(
A^{z=1}=1|v\right)  -\Pr\left(  A^{z=0}=1|v\right)  } \label{Marginal Wald}%
\end{equation}
\textit{where }$E\left(  Y^{z}|v\right)  =\int E\left(  Y|Z=z,L=l\right)
dF\left(  l|v\right)  $\textit{ and }$E\left(  A^{z}|v\right)  =\int E\left(
A|Z=z,L=l\right)  dF\left(  l|v\right)  .$

The result provides a causal interpretation of the nonparametric instrumental
variable estimand $\left(  \ref{Marginal Wald}\right)  $ as the NATE provided
$Z$ is a valid instrument for which the key condition $\left(  \ref{null cov}%
\right)  $ holds.\ Assumption $\left(  \ref{null cov}\right)  $ covers a range
of settings of potential practical interest; in fact at one end of the
identification spectrum covered by the assumption, suppose that the condition
holds for $V=\varnothing,$ that is $\mathrm{COV}\left(  \Delta_{y}\left(
U,L\right)  ,\pi\left(  U,L\right)  |\mathcal{N}=1\right)  =0,$ the result
then provides identification of the marginal NATE:
\[
\psi_{0}=E\left(  Y^{a=1}-Y^{a=0}|\mathcal{N}=1\right)  =\frac{E\left(
Y^{z=1}\right)  -E\left(  Y^{z=0}\right)  }{\Pr\left(  A^{z=1}=1\right)
-\Pr\left(  A^{z=0}=1\right)  }.
\]
At the other end of the spectrum where the condition holds for $V=L$; then the
result provides identification of $\psi\left(  L\right)  $ with the
corresponding conditional Wald ratio $\left(  \ref{Marginal Wald}\right)  $.
Interestingly, as further expanded on in the next Section, the stronger BCS
condition actually ensures stronger identification results as it guarantees
identification of any smooth functional of a given counterfactual outcome
among the nudge-able, and not only the conditional NATE; for instance, we show
that condition $\left(  \ref{BCS}\right)  $ provides identification of
$E\left(  Y^{a}|\mathcal{N}=1,L=l\right)  ,$ $a=0,1,$ while $\left(
\ref{null cov}\right)  $ only identifies the mean difference $E\left(
Y^{a=1}-Y^{a=0}|\mathcal{N}=1,L=l\right)  $ but not the separate
counterfactual means.

\section{The Generalized NATE\ Approach}

The previous Section focused primarily on the additive NATE, e.g. $E\left(
Y^{a=1}-Y^{a=0}|\mathcal{N}=1\right)  $; in the current Section we establish a
generalization of the result by providing identification of the conditional
cumulative distribution function of a counterfactual outcome $Y^{a}$ for the
subgroup of units whose treatment status is nudge-able , i.e. $\Pr\left(
Y^{a}\leq y|\mathcal{N}=1\right)  $, which implies nonparametric
identification of any smooth functional of the counterfactual distribution.
\ This general class of identified functionals includes any functional
potentially of interest in causal inference, provided that it would in
principle be identified under unconfoundedness. Our result provides
identification of common nonlinear counterfactual contrasts such as risk
ratios and odds ratios for binary outcomes, as well as causal contrasts for
censored time to event outcomes such as hazard ratios and hazard differences,
and quantile treatment effects for continuous outcomes. We state a general
form of the result which in principle allows for the counterfactual
distribution function of interest to condition on $V.$

\noindent\textit{Result 2: Suppose that conditions} $\left(
\ref{independence}\right)  ,\left(  \ref{Latent Unconfoundedness}\right)
,\left(  \ref{relevance}\right)  $\textit{ and the balanced compliers' share
condition, conditional on }$V:$%
\begin{equation}
\mathcal{C}\amalg\left(  U,W\right)  |\mathcal{N}=1,V \label{CICS}%
\end{equation}
\textit{hold, then we have that for any specified function }$h\left(
y,v\right)  ,$%

\begin{align}
\mu(a,v;h)  &  \mathrm{=}E\left(  h\left(  Y^{a},V\right)  |\mathcal{N}%
=1,V=v\right) \nonumber\\
&  =\frac{E\left(  \mathrm{I}\left(  A^{z=1}=a\right)  h\left(  Y^{z=1}%
,V\right)  |V=v\right)  -E\left(  \mathrm{I}\left(  A^{z=0}=a\right)  h\left(
Y^{z=0},V\right)  |V=v\right)  }{\Pr\left(  A^{z=1}=a|V=v\right)  -\Pr\left(
A^{z=0}=a|V=v\right)  } \label{Wald EE}%
\end{align}
\textit{where }%
\[
E\left(  \mathrm{I}\left(  A^{z}=a\right)  h\left(  Y^{z},V\right)
|V=v\right)  =\int E\left(  \mathrm{I}\left(  A=a\right)  h\left(  Y,V\right)
|Z=z,L=l\right)  dF\left(  l|V=v\right)
\]
\textit{ and}%
\[
\mathit{Pr}\left(  A^{z}=a|V=v\right)  =\int\Pr\left(  A=a|Z=z,L=l\right)
dF\left(  l|V=v\right)
\]
$.$

The result clearly generalizes Result 1 which is recovered by taking a
difference of $h\left(  Y^{a},V\right)  =Y^{a}$ for $a=0,1;$ furthermore, by
providing identification of the counterfactual mean $E(Y^{a}|\mathcal{N}%
=1,V=v),$ the result also accommodates other counterfactual contrasts$.$
Finally, an immediate consequence of this result is the opportunity it offers
to identify any counterfactual quantity that can potentially be expressed
implicitly as the solution to a moment equation. We briefly illustrate with
the example of the median nudge treatment effect of a continuous outcome, a
well-established nonlinear causal effect given by $\psi\left(  v\right)
=\overline{\mu}\left(  1,v\right)  -\overline{\mu}\left(  0,v\right)  $, where
$\overline{\mu}\left(  a,v\right)  \equiv F_{Y^{a}|\mathcal{N}=1,V=v}%
^{-1}\left(  1/2\right)  $ with $F_{Y^{a}|\mathcal{N}=1,V=v}^{-1}$ the
conditional quantile function of the counterfactual $Y^{a}$ given $\left\{
\mathcal{N}=1,V=v\right\}  .$ Then, under the conditions of Result 2, one can
show that $\overline{\mu}\left(  a,v\right)  $ is the unique solution to the
moment equation $E\left[  \left(  \left\{  I\left(  Y^{a}\leq\overline{\mu
}\left(  a,V\right)  \right)  -1/2\right\}  \right)  g\left(  V\right)
|\mathcal{N}=1\right]  =0,$ for any user-supplied function $g,$ which is
identified by equation $\left(  \ref{Wald EE}\right)  $ with $h\left(
y,v\right)  =\left\{  I\left(  y\leq\overline{\mu}\left(  a,v\right)  \right)
-1/2\right\}  g\left(  V\right)  $, $\ $leading to a feasible analytic
strategy to construct empirical moment equations as a basis for statistical
inference, whether parametric, semiparametric or nonparametric.

\section{A GLIM\ Generative IV Model}

In this section, we describe a generative treatment selection model in the
form of a generalized latent index model\ (GLIM) as a common platform to
relate NATE identifying assumptions to other IV\ models in the literature.
Throughout, to facilitate the exposition, measured covariates $L$ are
suppressed$.$ \ The GLIM\ specifies a selection model for the potential
treatment
\[
A^{z}=\mathrm{I}\left(  h\left(  z,U\right)  \geq\epsilon_{z}\right)  ,\text{
}z=0,1.
\]
for a specific class of functions $h\left(  z,U\right)  $ of the instrument,
and unmeasured confounders, and random threshold values $\left(
\epsilon_{z=0},\epsilon_{z=1}\right)  $ for each potential instrument value,
$\epsilon_{z}$ independent of $(Z,U).$ Notably, the GLIM\ does not impose any
restriction on the distribution of the potential outcomes $\{Y^{a}:a\}$ given
$U.$ Crucially, the GLIM\ includes as a special case, the LIM\ of Heckman
(1976), if (i) $h\left(  z,U\right)  =p\left(  z\right)  +U$; (ii)
$\epsilon_{z}$ is degenerate with all of its mass at 1,i.e. $\Pr\left\{
\epsilon_{z}=1\right\}  =1,$ $z=0,1;$ (iii) $U$ follows a uniform distribution
on the unit interval. \ According to a theorem due to Vytlacil (2002), the
LIM\ specification (i)-(iii) is equivalent to the LATE model of Imbens and
Angrist (1994) given by $\left(  \ref{IA Independence}\right)  $-$\left(
\ref{IA relevance}\right)  .$ To quote Vytlacil (2002): "the LATE\ conditions
and the LIM are not only indistinguishable based on observational data, but
they cannot be distinguished based on any hypothetical intervention or
experiment. The two models are equivalent." A key insight in the LIM
formulation of the LATE is that by setting $\epsilon_{z=0}=\epsilon_{z=1}=1,$
it effectively rules out any other cause for the treatment, beyond the IV and
scalar confounding factor $U!$

Interestingly, the following specification of a GLIM\ implies the additive
selection model of Wang and Tchetgen\ Tchetgen (2018) (i*) $h\left(
z,U\right)  =p\left(  z\right)  +U$ lies in the unit interval and
(ii*)$\epsilon_{z}$ uniformly distributed on the unit interval, $z=0,1;$ while
the following specification of a GLIM\ implies the multiplicative selection
model of Liu et al (2025) (i**)$h\left(  z,U\right)  =p\left(  z\right)
\times U$ lies in the unit interval and (ii**) $\epsilon_{z}$ uniform on the
unit interval, $z=0,1.$ We should note still that the latent unconfoundedness
conditions in the additive and multiplicative selection models differs
slightly, mainly the former imposes $A\amalg Y^{a}|U,a=0,1;$ while the latter
only requires $A\amalg Y^{a=0}|U.$ This should not be entirely surprising
given that the additive selection model identifies the population average
treatment effect, while the multiplicative selection model identifies the
average treatment effect for the treated only. One should also note that
presence of the stochastic thresholds $\epsilon_{z}$, $z=0,1$ in the additive
and multiplicative selection models generally rule out monotonicity, that is
whenever $\epsilon_{z=1}>h\left(  z=1,U\right)  >$ $h\left(  z=0,U\right)
>\epsilon_{z=0},$ such that $A^{z=1}=0,$ however $A^{z=0}=1,$ which can
clearly happen, even if $h\left(  z=1,U\right)  >h\left(  z=0,U\right)  $ for
all units$.$

Finally, we offer a specification of a GLIM\ which implies the
\textit{balanced compliers' share condition} $\mathcal{C}\amalg U|\mathcal{N}%
=1,$ and therefore constitutes a sufficient condition for identification of
the NATE. \ Let logit$\left(  x\right)  \equiv\log(x/(1-x)$ denote the logit
link function.

\bigskip

\smallskip

\textit{Result 3: Consider the GLIM\ with}$\mathit{\ }$i$^{\dag}$)$h\left(
z,U\right)  =p\left(  z\right)  +U;$\textit{ ii}$^{\dag}$\textit{)}%
$\epsilon_{z}$\textit{ follows a standard logistic distribution }%
\textrm{logistic(0,1)}\textit{\textrm{ }with distribution function given by}
\textrm{logit}$\Pr\left\{  \epsilon_{z}\leq e\right\}  =e,$\textit{ }%
$z=0,1;$\textit{ and }$\epsilon_{0}\amalg\epsilon_{1}$\textit{; then we have
that the BCS\ condition holds:}
\[
\mathcal{C}\amalg U|\mathcal{N}=1.
\]

The conditions i$^{\dag}$) and ii$^{\dag}$) imply the latent product binomial
propensity score model:
\begin{align*}
\text{logit}\Pr\left\{  A^{z}=1|U\right\}   &  =\text{logit}\Pr\left\{
A=1|Z=z,U\right\}  =p\left(  z\right)  +U;\\
&  A^{z=1}\amalg A^{z=0}|U;
\end{align*}
where the marginal logistic propensity score model encodes a no-odds ratio
interaction between the instrument and an unmeasured confounder, while the
conditional independence statement renders the potential treatments
independent, a relatively mild condition, given that $U$ is not required to be
observed. This model is akin to a standard logistic mixed effect model for
dependent binary variables.

Interestingly, the proposed GLIM\ justifying the BCS condition may also be
derived as a certain logit discrete choice model, a special case of a large
class of discrete choice models first introduced and extensively used in
economics and other social sciences, as a principled approach for generating
multinomial models to describe discrete choice decision making under rational
utility maximization (McFadden, 1984, Train, 2009). We refer to Klein (2010)
for a related discussion of the relationship between GLIMs and discrete choice
models. \ Briefly returning to the illustrative welfare program and Mendelian
randomization studies, the result provides a concrete and familiar latent
propensity score generative model for selection into the welfare program, or
obesity status as a function of the genetic instrument and hidden confounding
factors such that the BCS condition might be satisfied even though
monotonicity might not.

The GLIM platform is clarifying in that it provides a common framework by
which to articulate identification conditions of the LATE, the ATE, the ATT
and the NATE. It shows that identification of the LATE requires no
IV\ interaction with an unmeasured confounder on the additive scale,
\textit{and} that the latent threshold a unit's utility must cross for them to
take the treatment must remain constant across IV conditions; \textit{and}
that all units share a common utility under a given value of the instrument.
In contrast, no additive, no multiplicative, and no odds ratio interaction
between the IV and an unmeasured confounder at the utility level is a
sufficient (but not necessary) condition for the additive, multiplicative and
logistic selection model, to identify the ATE,\ ATT and NATE, respectively.
Crucially, these last three selection models accommodate both heterogeneous
thresholds and heterogeneous utility functions under a fixed value of the
instrument. Table 1. summarizes the GLIM model for each counterfactual mean
identified by the single-arm Wald ratio%
\[
\mu(a;y)=\frac{E\left(  \mathrm{I}\left(  A=a\right)  Y|Z=1\right)  -E\left(
\mathrm{I}\left(  A=a\right)  Y|Z=0\right)  }{\Pr\left(  A=a|Z=1\right)
-\Pr\left(  A=a|Z=0\right)  }%
\]

\bigskip

$\
\begin{tabular}
[c]{|l|l|l|}\hline
$\text{Identification Assumption}$ & $\text{GLIM Model Formulation}$ & Target
of Inference\\\hline
$\text{Monotonicity}$ & $%
\begin{array}
[c]{c}%
A^{z}=\mathrm{I}\left(  p\left(  z\right)  +U\geq\epsilon_{z}\right) \\
0\leq p\left(  0\right)  \leq p(1)\leq1\\
U\sim\text{Uniform}\left(  0,1\right)  ,\epsilon_{0}=\epsilon_{1}=1
\end{array}
$ & $%
\begin{array}
[c]{c}%
\mu(a;y)=E\left(  Y^{a}|\mathcal{C=}\mathrm{co}\right) \\
\lbrack\text{Imbens and Angrist(1994),}\\
\text{Vytlacil(2002)]}%
\end{array}
$\\\hline
$\text{Additive IV Model}$ & $%
\begin{array}
[c]{c}%
A^{z}=\mathrm{I}\left(  p\left(  z\right)  +U\geq\epsilon_{z}\right) \\
0\leq p\left(  z\right)  +U\leq1,\text{ and}\\
\epsilon_{z}\sim\text{Uniform}\left(  0,1\right)  ,z=0,1
\end{array}
$ & $%
\begin{array}
[c]{c}%
\mu(a;y)=E\left(  Y^{a}\right) \\
\lbrack\text{Wang and }\\
\text{Tchetgen Tchetgen, 2018]}%
\end{array}
$\\\hline
$\text{Multiplicative IV Model}$ & $%
\begin{array}
[c]{c}%
A^{z}=\mathrm{I}\left(  p\left(  z\right)  \cdot U\geq\epsilon_{z}\right) \\
0\leq p\left(  z\right)  \cdot U\leq1,\text{ and}\\
\epsilon_{z}\sim\text{Uniform}\left(  0,1\right)  ,z=0,1
\end{array}
$ & $%
\begin{array}
[c]{c}%
\mu(a=0;y)=E\left(  Y^{a=0}|A=1\right) \\
\lbrack\text{Liu et al, 2025, }\\
\text{and Lee et al, 2025]}%
\end{array}
$\\\hline
$\text{Logistic IV\ Model}$ & $%
\begin{array}
[c]{c}%
A^{z}=\mathrm{I}\left(  p\left(  z\right)  +U\geq\epsilon_{z}\right) \\
p\left(  z\right)  +U\in\lbrack-\infty,\infty]\\
\epsilon_{z}\sim\text{Logistic}\left(  0,1\right)  ,\epsilon_{0}\amalg
\epsilon_{1}%
\end{array}
$ & $%
\begin{array}
[c]{c}%
\mu(a;y)=E\left(  Y^{a}|\mathcal{N}=1\right) \\
\text{ This Work}%
\end{array}
$\\\hline
\end{tabular}
\ \ $

Table 1. IV\ Identification via GLIM Treatment Selection Models

\section{Estimation and Inference}

Under the proposed identifying conditions, estimation and inference for the
NATE (whether conditional or marginal with respect to $V),$ essentially boils
down to estimation and inference about a version of the Wald functional,
possibly conditional on covariates$;$ a statistical task for which several
existing statistical methods have already been developed which can readily be
adopted. For instance, considering the marginal NATE $\left(
\ref{Marginal Wald}\right)  $ of Result 1 where $V=\varnothing$, clearly the
goal reduces to estimating the two g-formulae $E\left(  Y^{z}\right)  =\int
E\left(  Y|Z=z,L=l\right)  dF\left(  l|v\right)  $\textit{ and }$E\left(
A^{z}\right)  =\int E\left(  A|Z=z,L=l\right)  dF\left(  l\right)  $, a task
for which several methods exist, including standard inverse probability
weighting, g-computation and doubly robust estimation following semiparametric
theory developed by James Robins and collaborators, see Hern\'{a}n and Robins
(2020) for an accessible textbook treatment of these methods and relevant
references for state of the art developments in this research area. Likewise,
several works have considered parametric and semiparametric methods for
targeting a model for the arm-specific Wald estimand $\left(  \ref{Wald EE}%
\right)  $ which can readily be applied to the setting of Result 2.
Application of these methods together with a standard use of the nonparametric
bootstrap readily provides appropriate measures of uncertainty including
well-calibrated large sample confidence intervals. Details are omitted,
although the reader may refer to Abadie (2003), Tan (2006) and Ogburn et al
(2015), Wang et al (2018) for a description of analogous methods that can
readily be adapted to the current setting.

\section{Discussion}

This paper contributes to a growing body of works on the causal interpretation
of the Wald ratio as a binary instrumental variable estimand (Robins, 1994,
Hern\'{a}n and Robins, 2006, Wang et al, 2018). A relevant prior work worth
mentioning in relation to our contribution is De Chaisemartin (2017) who
introduces the "compliers-defiers" assumption, which requires a subgroup of
compliers of the same size and with identical LATE as defiers, in which case
the Wald ratio identifies the LATE\ in the remaining compliers. \ Another
relevant paper, \ Small et al (2017) consider a relaxation of monotonicity
that only requires a monotonic increasing relationship to hold across subjects
(and not necessarily within subjects) between the IV and the treatments
conditionally on a set of unmeasured confounders. They show that under
stochastic monotonicity, the IV method identifies a weighted average of
treatment effects with greater weight on subgroups of subjects on whom the IV
has a stronger effect. We view these prior works and ours as separate and
complementary contributions to further advance our understanding of the
instrumental variable model when monotonicity may not hold. Interestingly, in
contrast with De Chaisemartin (2017) and similar to Small et al (2017), our
identifying assumptions are expressed explicitly in terms of the unobserved
confounder $U$, the most typical approach to reason about residual confounding
particularly in settings where the source of confounding may be known or can
at least be hypothesized on subject matter basis, but may not easily be
measured. In comparison, by suppressing $U$ as a potential source of
confounding De Chaisemartin (2017) offers a parsimonious set up where
reasoning about identifying conditions may be feasible without necessarily
having a specific hypothesis about the source of unmeasured confounding.
Another major difference between our works is that whereas De Chaisemartin
(2017) were primarily interested in identifying conditions under which the
Wald ratio preserves an interpretation as the average causal effect for a
subgroup of compliers, while Small et al (2017) identified the average causal
effect for a certain weighted population defined in terms of a tilted
distribution of the unmeasured confounders, with the tilt related to the
strength of the instrument effect on the treatment (as a function of $U$).
\ In contrast,\ our framework has instead focused on a different objective;
mainly on establishing conditions under which the Wald ratio may be
interpreted as the NATE. Finally, as we show, a stronger identifying condition
provides identification of any parameter of a treatment-specific potential
outcome marginal distribution among a given subgroup of nudge-able units on
any scale potentially of interest, for which the parameter would in principle
be identified under unconfoundedness.

We conclude with one final remark by noting that monotonicity not only
identifies the LATE, it also identifies the overall share of compliers in the
target population, i.e. $\Pr\left\{  \mathcal{C}=\mathrm{co}\right\}
=\Pr\left\{  A^{z=1}=1\right\}  -\Pr\left\{  A^{z=0}=1\right\}  ,$ which
yields $\Pr\left\{  \mathcal{C}=\mathrm{co}\right\}  =\Pr\left\{
A=1|Z=1\right\}  -\Pr\left\{  A=1|Z=0\right\}  \equiv q$ is case of a
randomized instrument; although monotonicity does not identify the actual
subset of compliers in the population. In contrast, neither the share of
compliers nor that of defiers among nudge-able units (or in the underlying
population) appears to be identified under our conditions without a further
assumption. \ This should not come as a surprise, since afterall, a person's
compliance status involves the joint potential treatments $\{A^{z=1}%
,A^{z=0}\}$ which are inherently unobserved. Monotonicity attains
identification of the share of compliers by rendering the joint potential
treatments distribution degenerate, a strong restriction we avoid.
Nevertheless, as established in the next result, it remains possible to obtain
bounds for the share of compliers, defiers and thus for the overall share of
nudge-able units. Specifically, without any additional assumption besides
having access to a valid instrument, a straightforward application of
Fr\'{e}chet-Hoeffding bounds (Nelsen, 2006) gives
\begin{align*}
&  \max\left(  0,\Pr\left\{  A^{z=0}=1\mathrm{|}V=v\right\}  -\Pr\left\{
A^{z=1}=1\mathrm{|}V=v\right\}  \right) \\
&  \leq\Pr\left\{  \mathcal{C}=\mathrm{de|}V=v\right\} \\
&  \leq\min\left(  \Pr\left\{  A^{z=1}=0\mathrm{|}V=v\right\}  ,\Pr\left\{
A^{z=0}=1\mathrm{|}V=v\right\}  \right)  ;
\end{align*}
and
\begin{align*}
&  \max\left(  0,\Pr\left\{  A^{z=1}=1\mathrm{|}V=v\right\}  -\Pr\left\{
A^{z=0}=1\mathrm{|}V=v\right\}  \right) \\
&  \leq\Pr\left\{  \mathcal{C}=\mathrm{co|}V=v\right\} \\
&  \leq\min\left(  \Pr\left\{  A^{z=1}=1\mathrm{|}V=v\right\}  ,\Pr\left\{
A^{z=0}=0\mathrm{|}V=v\right\}  \right)  ,
\end{align*}
where $\Pr\left\{  A^{z}=a\mathrm{|}V=v\right\}  ,$ $z,a=0,1$ is identified
empirically by the standard g-formula. We recommend that these bounds be
routinely reported in practice. Specifically, to illustrate, in absence of
covariates and assuming that the instrument $Z$ is randomized and that it
generally encourages treatment uptake, such that $\Pr\left\{  A=1|Z=1\right\}
\geq\Pr\left\{  A=1|Z=0\right\}  $, we have that $0\leq\Pr\left\{
\mathcal{C}=\mathrm{de}\right\}  \leq\min\left(  \Pr\left\{  A=0\mathrm{|}%
Z=1\right\}  ,\Pr\left\{  A=1|Z=0\right\}  \right)  ,$ i.e. the proportion of
defiers cannot exceed the proportion of individuals who fail to adhere to
their assigned treatment in each arm. Crucially, these bounds accommodate the
special case of monotonicity, i.e. $\Pr\left\{  \mathcal{C}=\mathrm{de}%
\right\}  =0,$ which is nested within our model. Likewise, we have that
$\Pr\left\{  A=1\mathrm{|}Z=1\right\}  -\Pr\left\{  A=1\mathrm{|}Z=0\right\}
\leq\Pr\left\{  \mathcal{C}=\mathrm{co}\right\}  \leq\min\left(  \Pr\left\{
A=1\mathrm{|}Z=1\right\}  ,\Pr\left\{  A=0\mathrm{|}Z=0\right\}  \right)  ,$
such that the share of compliers is no smaller than $q,$ i.e. what it would be
under monotonicity, and it cannot exceed the proportion of individuals who
adhere to their assigned treatment in each arm; these are in effect, their
natural bounds.

\newpage

\begin{center}
{\Large REFERENCES}
\end{center}

\bigskip

Abadie, A., 2003. Semiparametric instrumental variable estimation of treatment
response models. Journal of econometrics, 113(2), pp.231-263.

Imbens GW, Angrist JD. Identification and Estimation of Local Average
Treatment Effects. Econometrica. 1994 Mar;62(2):467-75.

Baker, S.G. and Lindeman, K.S., 1994. The paired availability design: a
proposal for evaluating epidural analgesia during labor. Statistics in
Medicine, 13(21), pp.2269-2278.

Boef AG, le Cessie S, Dekkers OM, Frey P, Kearney PM, Kerse N, Mallen CD,
McCarthy VJ, Mooijaart SP, Muth C, Rodondi N. Physician's prescribing
preference as an instrumental variable: exploring assumptions using survey
data. Epidemiology. 2016 Mar 1;27(2):276-83.

De Chaisemartin, C., 2017. Tolerating defiance? Local average treatment
effects without monotonicity. Quantitative Economics, 8(2), pp.367-396.

Hern\'{a}n, M.A. and Robins, J.M., 2006. Instruments for causal inference: an
epidemiologist's dream?. Epidemiology, 17(4), pp.360-372.

Hern\'{a}n MA, Robins JM (2020). Causal Inference: What If. Boca Raton:
Chapman \& Hall/CRC.

Klein, T.J., 2010. Heterogeneous treatment effects: Instrumental variables
without monotonicity?. Journal of Econometrics, 155(2), pp.99-116.

Lee, Y., Yu, M., Liu, J., Park, C., Zhang, Y., Robins, J.M. and Tchetgen
Tchetgen, E.J., 2025. Inference on Nonlinear Counterfactual Functionals under
a Multiplicative IV Model. arXiv preprint arXiv:2507.15612.

Liu, J., Park, C., Lee, Y., Zhang, Y., Yu, M., Robins, J.M. and Tchetgen
Tchetgen, E.J., 2025. The Multiplicative Instrumental Variable Model. arXiv
preprint arXiv:2507.09302.

McFadden, D. L. (1984). Econometric analysis of qualitative response models.
Handbook of Econometrics, Volume II. Chapter 24. Elsevier Science Publishers BV.

Nelsen, R. B. (2006). An Introduction to Copulas, 2nd ed. Springer, New York.

Ogburn, E.L., Rotnitzky, A. and Robins, J.M., 2015. Doubly robust estimation
of the local average treatment effect curve. Journal of the Royal Statistical
Society Series B: Statistical Methodology, 77(2), pp.373-396.

Permutt, T. and Hebel, J.R., 1989. Simultaneous-equation estimation in a
clinical trial of the effect of smoking on birth weight. Biometrics, pp.619-622.

Robins, J.M., 1986. A new approach to causal inference in mortality studies
with a sustained exposure period---application to control of the healthy
worker survivor effect. Mathematical modelling, 7(9-12), pp.1393-1512.

Robins, J.M., 1994. Correcting for non-compliance in randomized trials using
structural nested mean models. Communications in Statistics-Theory and
methods, 23(8), pp.2379-2412.

Small, D.S., Tan, Z., Ramsahai, R.R., Lorch, S.A. and Brookhart, M.A., 2017.
Instrumental variable estimation with a stochastic monotonicity assumption.

Swanson SA, Miller M, Robins JM, Hern\'{a}n MA. Definition and evaluation of
the monotonicity condition for preference-based instruments. Epidemiology.
2015 May 1;26(3):414-20.

Tan, Z., 2006. Regression and weighting methods for causal inference using
instrumental variables. Journal of the American Statistical Association,
101(476), pp.1607-1618.

Train, K. (2009). Discrete Choice Methods with Simulation. Cambridge
University Press.

Wang, L. and Tchetgen Tchetgen, E., 2018. Bounded, efficient and multiply
robust estimation of average treatment effects using instrumental variables.
Journal of the Royal Statistical Society Series B: Statistical Methodology,
80(3), pp.531-550.

\newpage

\noindent{\Large APPENDIX}

\underline{Proof of Result 1:} Note that
\begin{align*}
&  E\left(  Y^{z=1}|V\right)  -E\left(  Y^{z=0}|V\right) \\
&  =E\left(  \left(  Y^{a=1}-Y^{a=0}\right)  A^{z=1}+Y^{a=0}|V\right)  \text{
}\\
&  -E\left(  \left(  Y^{a=1}-Y^{a=0}\right)  A^{z=0}+Y^{a=0}|V\right)  \text{
by consistency and exclusion restriction }\\
&  =E\left(  \left(  Y^{a=1}-Y^{a=0}\right)  \left(  A^{z=1}-A^{z=0}\right)
|V\right) \\
&  =E\left(  \left(  Y^{a=1}-Y^{a=0}\right)  \left(  \mathrm{I}\left\{
A^{z=1}>A^{z=0}\right\}  -\mathrm{I}\left\{  A^{z=1}<A^{z=0}\right\}  \right)
|V\right) \\
&  =E\left(  E\left[  \left(  Y^{a=1}-Y^{a=0}\right)  \left(  \mathrm{I}%
\left\{  \mathcal{C=}\mathrm{co}\right\}  -\mathrm{I}\left\{  \mathcal{C=}%
\mathrm{de}\right\}  \right)  |U,L\right]  |V\right)  \text{ by iterated
expectations}\\
&  =E\left(  \left.
\begin{array}
[c]{c}%
E\left(  Y^{a=1}-Y^{a=0}|U,L\right) \\
\times\left(  \Pr\left\{  \mathcal{C=}\mathrm{co}|U,L\right\}  -\mathrm{\Pr
}\left\{  \mathcal{C=}\mathrm{de}|U,L\right\}  \right)
\end{array}
\right\vert V\right)  \text{ by latent unconfoundedness }\\
&  =E\left(
\begin{array}
[c]{c}%
E\left(  Y^{a=1}-Y^{a=0}|U,L\right)  \left(  \frac{\Pr\left\{  \mathcal{C=}%
\mathrm{co}|U,L\right\}  -\mathrm{\Pr}\left\{  \mathcal{C=}\mathrm{de}%
|U,L\right\}  }{\Pr\left\{  \mathcal{C=}\mathrm{co}|U,L\right\}  +\mathrm{\Pr
}\left\{  \mathcal{C=}\mathrm{de}|U,L\right\}  }\right) \\
\times\left(  \Pr\left\{  \mathcal{C=}\mathrm{co}|U,L\right\}  +\mathrm{\Pr
}\left\{  \mathcal{C=}\mathrm{de}|U,L\right\}  \right)  |V
\end{array}
\right) \\
&  =E\left(  E\left(  Y^{a=1}-Y^{a=0}|U,L\right)  \left(
\begin{array}
[c]{c}%
\Pr\left\{  \mathcal{C=}\mathrm{co}|\mathcal{N}=1,U,L\right\} \\
-\Pr\left\{  \mathcal{C=}\mathrm{de}|\mathcal{N}=1,U,L\right\}
\end{array}
\right)  \Pr\left\{  \mathcal{N}=1|U,L\right\}  |V\right) \\
&  =E\left(  E\left(  Y^{a=1}-Y^{a=0}|\mathcal{N}=1,U,L\right)  \left(
2\Pr\left\{  \mathcal{C=}\mathrm{co}|\mathcal{N}=1,U,L\right\}  -1\right)
\Pr\left\{  \mathcal{N}=1|U,L\right\}  |V\right) \\
&  =E\left(  \left[
\begin{array}
[c]{c}%
E\left(  Y^{a=1}-Y^{a=0}|\mathcal{N}=1,U,L\right) \\
-E\left(  Y^{a=1}-Y^{a=0}|\mathcal{N}=1,V\right)
\end{array}
\right]  \left(
\begin{array}
[c]{c}%
2\Pr\left\{  \mathcal{C=}\mathrm{co}|\mathcal{N}=1,U,L\right\} \\
-1
\end{array}
\right)  \Pr\left\{  \mathcal{N}=1|U,L\right\}  |V\right) \\
&  +E\left(  \left[  E\left(  Y^{a=1}-Y^{a=0}|\mathcal{N}=1,V\right)  \right]
\left(  2\Pr\left\{  \mathcal{C=}\mathrm{co}|\mathcal{N}=1,U,L\right\}
-1\right)  \Pr\left\{  \mathcal{N}=1|U,L\right\}  |V\right) \\
&  =E\left(  \left[
\begin{array}
[c]{c}%
E\left(  Y^{a=1}-Y^{a=0}|\mathcal{N}=1,U,L\right) \\
-E\left(  Y^{a=1}-Y^{a=0}|\mathcal{N}=1,V\right)
\end{array}
\right]  \left(
\begin{array}
[c]{c}%
2\Pr\left\{  \mathcal{C=}\mathrm{co}|\mathcal{N}=1,U,L\right\} \\
-1
\end{array}
\right)  |V,\mathcal{N}=1\right)  \Pr\left\{  \mathcal{N}=1|V\right\} \\
&  +E\left(  \left[  E\left(  Y^{a=1}-Y^{a=0}|\mathcal{N}=1,V\right)  \right]
\left(
\begin{array}
[c]{c}%
2\Pr\left\{  \mathcal{C=}\mathrm{co}|\mathcal{N}=1,U,L\right\} \\
-1
\end{array}
\right)  |V,\mathcal{N}=1\right)  \Pr\left\{  \mathcal{N}=1|V\right\} \\
&  \underset{=2\cdot\mathrm{COV}\left(  E\left(  Y^{a=1}-Y^{a=0}%
|\mathcal{N}=1,U,L\right)  ,\Pr\left\{  \mathcal{C=}\mathrm{co}|\mathcal{N}%
=1,U,L\right\}  |\mathcal{N}=1,V\right)  }{=\underbrace{E\left(  \left[
\begin{array}
[c]{c}%
E\left(  Y^{a=1}-Y^{a=0}|\mathcal{N}=1,U,L\right) \\
-E\left(  Y^{a=1}-Y^{a=0}|\mathcal{N}=1,V\right)
\end{array}
\right]  \left(
\begin{array}
[c]{c}%
2\Pr\left\{  \mathcal{C=}\mathrm{co}|\mathcal{N}=1,U,L\right\} \\
-2\Pr\left\{  \mathcal{C=}\mathrm{co}|\mathcal{N}=1,V\right\}
\end{array}
\right)  |\mathcal{N}=1,V\right)  }}\Pr\left\{  \mathcal{N}=1V\right\} \\
&  +\underset{=0}{\underbrace{E\left(  \left[
\begin{array}
[c]{c}%
E\left(  Y^{a=1}-Y^{a=0}|\mathcal{N}=1,U,V\right) \\
-E\left(  Y^{a=1}-Y^{a=0}|\mathcal{N}=1,V\right)
\end{array}
\right]  \left(  2\Pr\left\{  \mathcal{C=}\mathrm{co}|\mathcal{N}=1,V\right\}
-1\right)  |\mathcal{N}=1,V\right)  \Pr\left\{  \mathcal{N}=1|V\right\}  }}\\
&  +\underset{=0}{\underbrace{E\left(  \left[  E\left(  Y^{a=1}-Y^{a=0}%
|\mathcal{N}=1,V\right)  \right]  \left(
\begin{array}
[c]{c}%
2\Pr\left\{  \mathcal{C=}\mathrm{co}|\mathcal{N}=1,U,V\right\} \\
-2\Pr\left\{  \mathcal{C=}\mathrm{co}|\mathcal{N}=1,V\right\}
\end{array}
\right)  |\mathcal{N}=1,V\right)  \Pr\left\{  \mathcal{N}=1|V\right\}  }}\\
&  +E\left(  Y^{a=1}-Y^{a=0}|\mathcal{N}=1,V\right)  \left(  2\Pr\left\{
\mathcal{C=}\mathrm{co}|\mathcal{N}=1,V\right\}  -1\right)  \Pr\left\{
\mathcal{N}=1|V\right\} \\
&  =2\cdot\mathrm{COV}\left(  E\left(  Y^{a=1}-Y^{a=0}|\mathcal{N}%
=1,U,L\right)  ,\Pr\left\{  \mathcal{C=}\mathrm{co}|\mathcal{N}=1,U,L\right\}
|\mathcal{N}=1,V\right)  \Pr\left\{  \mathcal{N}=1|V\right\} \\
&  +E\left(  Y^{a=1}-Y^{a=0}|\mathcal{N}=1,V\right)  \left(  2\Pr\left\{
\mathcal{C=}\mathrm{co}|\mathcal{N}=1,V\right\}  -1\right)  \Pr\left\{
\mathcal{N}=1|V\right\} \\
&  =2\cdot\mathrm{COV}\left(  E\left(  Y^{a=1}-Y^{a=0}|\mathcal{N}%
=1,U,L\right)  ,\Pr\left\{  \mathcal{C=}\mathrm{co}|\mathcal{N}=1,U,L\right\}
|\mathcal{N}=1,V\right)  \Pr\left\{  \mathcal{N}=1|V\right\} \\
&  +E\left(  Y^{a=1}-Y^{a=0}|\mathcal{N}=1,V\right)  \left(  \Pr\left\{
\mathcal{C=}\mathrm{co}|\mathcal{N}=1,V\right\}  -\Pr\left\{  \mathcal{C=}%
\mathrm{de}|\mathcal{N}=1,V\right\}  \right)  \Pr\left\{  \mathcal{N}%
=1|V\right\} \\
&  =2\cdot\underset{=0}{\underbrace{\mathrm{COV}\left(  E\left(
Y^{a=1}-Y^{a=0}|\mathcal{N}=1,U,L\right)  ,\Pr\left\{  \mathcal{C=}%
\mathrm{co}|\mathcal{N}=1,U,L\right\}  |\mathcal{N}=1,V\right)  }}\Pr\left\{
\mathcal{N}=1|V\right\} \\
&  +\left[  E\left(  Y^{a=1}-Y^{a=0}|\mathcal{N}=1,V\right)  \right]  \left(
\Pr\left\{  \mathcal{C=}\mathrm{co}|V\right\}  -\Pr\left\{  \mathcal{C=}%
\mathrm{de}|V\right\}  \right)
\end{align*}
therefore we conclude that
\[
E\left(  Y^{z=1}|V\right)  -E\left(  Y^{z=0}|V\right)  =E\left(
Y^{a=1}-Y^{a=0}|\mathcal{N}=1,V\right)  \left(  \Pr\left\{  \mathcal{C=}%
\mathrm{co}|V\right\}  -\Pr\left\{  \mathcal{C=}\mathrm{de}|V\right\}
\right)
\]
provided that
\[
\mathrm{COV}\left(  E\left(  Y^{a=1}-Y^{a=0}|\mathcal{N}=1,U,L\right)
,\Pr\left\{  \mathcal{C=}\mathrm{co}|\mathcal{N}=1,U,L\right\}  |\mathcal{N}%
=1,V\right)  =0
\]
and therefore
\begin{align*}
E\left(  Y^{a=1}-Y^{a=0}|\mathcal{N}=1,V=v\right)   &  =\frac{E\left(
Y^{z=1}|V=v\right)  -E\left(  Y^{z=0}|V=v\right)  }{\left(  \Pr\left\{
\mathcal{C=}\mathrm{co}|V=v\right\}  -\Pr\left\{  \mathcal{C=}\mathrm{de}%
|V=v\right\}  \right)  }\\
&  =\frac{E\left(  Y^{z=1}|V=v\right)  -E\left(  Y^{z=0}|V=v\right)  }{\left(
\Pr\left\{  A^{z=1}=1|V=v\right\}  -\Pr\left\{  A^{z=0}=1|V=v\right\}
\right)  }%
\end{align*}
proving the result.

\underline{Proof of Result 2}:

\bigskip\ Note that by an analogous argument as above,%

\begin{align*}
&  E\left(  \mathrm{I}\left(  A^{z=1}=a\right)  h\left(  Y^{z=1},V\right)
|V=v\right)  -E\left(  \mathrm{I}\left(  A^{z=0}=a\right)  h\left(
Y^{z=0},V\right)  |V=v\right) \\
&  =E\left(  \mathrm{I}\left(  A^{z=1}=a\right)  h\left(  Y^{a},V\right)
|V=v\right)  -E\left(  \mathrm{I}\left(  A^{z=0}=a\right)  h\left(
Y^{a},V\right)  |V=v\right) \\
&  =E\left(  \left[  \mathrm{I}\left(  A^{z=1}=a\right)  -\mathrm{I}\left(
A^{z=0}=a\right)  \right]  h\left(  Y^{a},V\right)  |V=v\right) \\
&  =E\left(  \left[  \mathrm{I}\left(  A^{z=1}=a,A^{z=0}=1-a\right)
-\mathrm{I}\left(  A^{z=1}=1-a,A^{z=0}=a\right)  \right]  h\left(
Y^{a},V\right)  |V=v\right) \\
&  =E\left(  E\left[  \mathrm{I}\left(  A^{z=1}=a,A^{z=0}=1-a\right)
-\mathrm{I}\left(  A^{z=1}=1-a,A^{z=0}=a\right)  |U,L\right]  h\left(
Y^{a},V\right)  |V=v\right)
\end{align*}
suppose $a=1$ then
\begin{align*}
&  E\left(  \mathrm{I}\left(  A^{z=1}=a\right)  h\left(  Y^{z=1},V\right)
|V=v\right)  -E\left(  \mathrm{I}\left(  A^{z=0}=a\right)  h\left(
Y^{z=0},V\right)  |V=v\right) \\
&  =E\left(  E\left[  \mathrm{I}\left(  A^{z=1}=1,A^{z=0}=0\right)
-\mathrm{I}\left(  A^{z=1}=0,A^{z=0}=1\right)  |U,L\right]  h\left(
Y^{a=1},V\right)  |V=v\right)  \text{ by exclusion restriction}\\
&  =E\left(  \left[  \Pr\left(  A^{z=1}>A^{z=0}\ |U,L\right)  -\mathrm{\Pr
}\left(  A^{z=1}<A^{z=0}|U,L\right)  \right]  h\left(  Y^{a=1},V\right)
|V=v\right)  \text{ by iterated expectations}\\
&  =E\left(  \frac{\left[  \Pr\left(  A^{z=1}>A^{z=0}\ |U,L\right)
-\mathrm{\Pr}\left(  A^{z=1}<A^{z=0}|U,L\right)  \right]  }{\left[  \Pr\left(
A^{z=1}>A^{z=0}\ |U,L\right)  +\mathrm{\Pr}\left(  A^{z=1}<A^{z=0}|U,L\right)
\right]  }\left[
\begin{array}
[c]{c}%
\Pr\left(  A^{z=1}>A^{z=0}|U,L\right) \\
+\mathrm{\Pr}\left(  A^{z=1}<A^{z=0}|U,L\right)
\end{array}
\right]  h\left(  Y^{a=1},V\right)  |V=v\right) \\
&  =E\left(  \ \left(  \frac{\Pr\left\{  \mathcal{C=}\mathrm{co}|U,L\right\}
-\mathrm{\Pr}\left\{  \mathcal{C=}\mathrm{de}|U,L\right\}  }{\Pr\left\{
\mathcal{C=}\mathrm{co}|U,L\right\}  +\mathrm{\Pr}\left\{  \mathcal{C=}%
\mathrm{de}|U,L\right\}  }\right)  \left[
\begin{array}
[c]{c}%
\Pr\left\{  \mathcal{C=}\mathrm{co}|U,L\right\} \\
+\mathrm{\Pr}\left\{  \mathcal{C=}\mathrm{de}|U,L\right\}
\end{array}
\right]  h\left(  Y^{a=1},V\right)  |V=v\right) \\
&  =E\left(  \ \left(  \Pr\left\{  \mathcal{C=}\mathrm{co}|\mathcal{N}%
=1,U,L\right\}  -\Pr\left\{  \mathcal{C=}\mathrm{de}|\mathcal{N}%
=1,U,L\right\}  \right)  \Pr\left\{  \mathcal{N}=1|U,L\right\}  h\left(
Y^{a=1},V\right)  |V=v\right) \\
&  =E\left(  \ \left(  2\Pr\left\{  \mathcal{C=}\mathrm{co}|\mathcal{N}%
=1,U,L\right\}  -1\right)  \Pr\left\{  \mathcal{N}=1|U,L\right\}  h\left(
Y^{a=1},V\right)  |V=v\right) \\
&  =E\left(  \ \left(  2\Pr\left\{  \mathcal{C=}\mathrm{co}|\mathcal{N}%
=1,V\right\}  -1\right)  \Pr\left\{  \mathcal{N}=1|U,L\right\}  h\left(
Y^{a=1},V\right)  |V=v\right)  \text{ by BCS}\\
&  =\left(  2\Pr\left\{  \mathcal{C=}\mathrm{co}|\mathcal{N}=1,V\right\}
-1\right)  \Pr\left\{  \mathcal{N}=1|V\right\}  E\left[  h\left(
Y^{a=1},V\right)  |\mathcal{N}=1,V\right]  \text{ by iterated expectations}\\
&  =\ \left(  \Pr\left\{  \mathcal{C=}\mathrm{co}|V\right\}  -\mathrm{\Pr
}\left\{  \mathcal{C=}\mathrm{de}|V\right\}  \right)  E\left[  h\left(
Y^{a=1},V\right)  |\mathcal{N}=1,V\right] \\
&  =\left(  \Pr\left\{  A^{z=1}=1|V=v\right\}  -\Pr\left\{  A^{z=0}%
=1|V=v\right\}  \right)  E\left[  h\left(  Y^{a=1},V\right)  |\mathcal{N}%
=1,V\right]
\end{align*}
Likewise, for $a=0,$%
\begin{align*}
&  E\left(  \mathrm{I}\left(  A^{z=1}=0\right)  h\left(  Y^{z=1},V\right)
|V=v\right)  -E\left(  \mathrm{I}\left(  A^{z=0}=0\right)  h\left(
Y^{z=0},V\right)  |V=v\right) \\
&  =E\left(  E\left[  \mathrm{I}\left(  A^{z=1}=0,A^{z=0}=1\right)
-\mathrm{I}\left(  A^{z=1}=1,A^{z=0}=0\right)  |U,L\right]  h\left(
Y^{a=0},V\right)  |V=v\right) \\
&  =E\left(  \left[  \Pr\left(  A^{z=0}>A^{z=1}\ |U,L\right)  -\mathrm{\Pr
}\left(  A^{z=0}<A^{z=1}|U,L\right)  \right]  h\left(  Y^{a=0},V\right)
|V=v\right) \\
&  =E\left(  \frac{\left[  \Pr\left(  A^{z=1}<A^{z=0}\ |U,L\right)
-\mathrm{\Pr}\left(  A^{z=1}>A^{z=0}|U,L\right)  \right]  }{\left[  \Pr\left(
A^{z=1}>A^{z=0}\ |U,L\right)  +\mathrm{\Pr}\left(  A^{z=1}<A^{z=0}|U,L\right)
\right]  }\left[
\begin{array}
[c]{c}%
\Pr\left(  A^{z=1}<A^{z=0}\ |U,L\right) \\
+\mathrm{\Pr}\left(  A^{z=1}>A^{z=0}|U,L\right)
\end{array}
\right]  h\left(  Y^{a=0},V\right)  |V=v\right) \\
&  =E\left(  \ \left(  \frac{\Pr\left\{  \mathcal{C=}\mathrm{de}|U,L\right\}
-\mathrm{\Pr}\left\{  \mathcal{C=}\mathrm{co}|U,L\right\}  }{\Pr\left\{
\mathcal{C=}\mathrm{co}|U,L\right\}  +\mathrm{\Pr}\left\{  \mathcal{C=}%
\mathrm{de}|U,L\right\}  }\right)  \left[
\begin{array}
[c]{c}%
\Pr\left\{  \mathcal{C=}\mathrm{co}|U,L\right\} \\
+\mathrm{\Pr}\left\{  \mathcal{C=}\mathrm{de}|U,L\right\}
\end{array}
\right]  h\left(  Y^{a=0},V\right)  |V=v\right) \\
&  =E\left(  \ \left(  \Pr\left\{  \mathcal{C=}\mathrm{de}|\mathcal{N}%
=1,U,L\right\}  -\Pr\left\{  \mathcal{C=}\mathrm{co}|\mathcal{N}%
=1,U,L\right\}  \right)  \Pr\left\{  \mathcal{N}=1|U,L\right\}  h\left(
Y^{a=0},V\right)  |V=v\right) \\
&  =E\left(  \ \left(  2\Pr\left\{  \mathcal{C=}\mathrm{de}|\mathcal{N}%
=1,U,L\right\}  -1\right)  \Pr\left\{  \mathcal{N}=1|U,L\right\}  h\left(
Y^{a=0},V\right)  |V=v\right) \\
&  =E\left(  \ \left(  2\Pr\left\{  \mathcal{C=}\mathrm{de}|\mathcal{N}%
=1,V\right\}  -1\right)  \Pr\left\{  \mathcal{N}=1|U,L\right\}  h\left(
Y^{a=0},V\right)  |V=v\right) \\
&  =\left(  2\Pr\left\{  \mathcal{C=}\mathrm{de}|\mathcal{N}=1,V\right\}
-1\right)  \Pr\left\{  \mathcal{N}=1|V\right\}  E\left[  h\left(
Y^{a=0},V\right)  |\mathcal{N}=1,V=v\right] \\
&  =\ \left(  \Pr\left\{  \mathcal{C=}\mathrm{de}|V\right\}  -\mathrm{\Pr
}\left\{  \mathcal{C=}\mathrm{co}|V\right\}  \right)  E\left[  h\left(
Y^{a=0},V\right)  |\mathcal{N}=1,V=v\right] \\
&  =\left(  \Pr\left(  A^{z=0}>A^{z=1}\ |V\right)  -\mathrm{\Pr}\left(
A^{z=0}<A^{z=1}|V\right)  \right)  E\left[  h\left(  Y^{a=0},V\right)
|\mathcal{N}=1,V=v\right] \\
&  =\left(  E\left[  \mathrm{I}\left(  A^{z=1}=0,A^{z=0}=1\right)
-\mathrm{I}\left(  A^{z=1}=1,A^{z=0}=0\right)  |V\right]  \right)  E\left[
h\left(  Y^{a=0},V\right)  |\mathcal{N}=1,V=v\right] \\
&  =\left(  E\left[  \mathrm{I}\left(  A^{z=1}=0\right)  -\mathrm{I}\left(
A^{z=0}=0\right)  |V\right]  \right)  E\left[  h\left(  Y^{a=0},V\right)
|\mathcal{N}=1,V=v\right] \\
&  =\left(  \Pr\left(  A^{z=1}=0|V\right)  -\Pr\left(  A^{z=0}=0|V\right)
\right)  E\left[  h\left(  Y^{a=0},V\right)  |\mathcal{N}=1,V=v\right]
\end{align*}
Therefore,
\begin{align*}
&  E\left(  \mathrm{I}\left(  A^{z=1}=a\right)  h\left(  Y^{z=1},V\right)
|V=v\right)  -E\left(  \mathrm{I}\left(  A^{z=0}=a\right)  h\left(
Y^{z=0},V\right)  |V=v\right) \\
&  =\left(  \Pr\left\{  A^{z=1}=1|V=v\right\}  -\Pr\left\{  A^{z=0}%
=1|V=v\right\}  \right)  E\left[  h\left(  Y^{a=1},V\right)  |\mathcal{N}%
=1,V=v\right] \\
&  I\left(  a=1\right) \\
&  +\left(  \Pr\left(  A^{z=1}=0|V\right)  -\Pr\left(  A^{z=0}=0|V\right)
\right)  E\left[  h\left(  Y^{a=0},V\right)  |\mathcal{N}=1,V=v\right] \\
&  I\left(  a=0\right) \\
&  =\left(  \Pr\left\{  A^{z=1}=a|V=v\right\}  -\Pr\left\{  A^{z=0}%
=a|V=v\right\}  \right)  E\left[  h\left(  Y^{a},V\right)  |\mathcal{N}%
=1,V=v\right]
\end{align*}
such that
\begin{align*}
\mu(a,v;h)  &  \mathrm{=}E\left(  h\left(  Y^{a},V\right)  |\mathcal{N}%
=1,V=v\right) \\
&  =\frac{E\left(  \mathrm{I}\left(  A^{z=1}=a\right)  h\left(  Y^{z=1}%
,V\right)  |V=v\right)  -E\left(  \mathrm{I}\left(  A^{z=0}=a\right)  h\left(
Y^{z=0},V\right)  |V=v\right)  }{\Pr\left(  A^{z=1}=a|V=v\right)  -\Pr\left(
A^{z=0}=a|V=v\right)  }%
\end{align*}
proving the result.

\bigskip

\underline{Proof of Result 3}:

\bigskip\textit{ }The conditions i$^{\dag}$) and ii$^{\dag}$) imply the latent
propensity score model
\begin{align*}
&  \text{logit}\Pr\left\{  A_{z}=1|U\right\} \\
&  =\text{logit}\Pr\left\{  A=1|Z=z,U\right\} \\
&  =\text{logit}\mathrm{\Pr}\left(  h\left(  Z,U\right)  \geq\epsilon
_{z}|Z=z,U\right) \\
&  =\text{logit}\left(  \text{expit}\left\{  p\left(  z\right)  +U\right\}
\right)  =p\left(  z\right)  +U
\end{align*}
furthermore, ii$^{\dag}$) implies
\begin{align*}
&  \Pr\left\{  A^{z=0}=a_{0},A^{z=1}=a_{1}|U\right\} \\
&  =\Pr\left\{  h\left(  0,U\right)  \leq\epsilon_{0},h\left(  1,U\right)
\leq\epsilon_{1}|U\right\}  ^{\mathrm{I}\left(  a_{0}=1,\mathrm{a}%
_{1}=1\right)  }\\
&  \times\Pr\left\{  h\left(  0,U\right)  \leq\epsilon_{0},h\left(
1,U\right)  >\epsilon_{1}|U\right\}  ^{\mathrm{I}\left(  a_{0}=1,\mathrm{a}%
_{1}=0\right)  }\\
&  \times\Pr\left\{  h\left(  0,U\right)  >\epsilon_{0},h\left(  1,U\right)
\leq\epsilon_{1}|U\right\}  ^{\mathrm{I}\left(  a_{0}=0,\mathrm{a}%
_{1}=1\right)  }\\
&  \times\Pr\left\{  h\left(  0,U\right)  >\epsilon_{0},h\left(  1,U\right)
>\epsilon_{1}|U\right\}  ^{\mathrm{I}\left(  a_{0}=0,\mathrm{a}_{1}=0\right)
}\\
&  =\Pr\left\{  h\left(  0,U\right)  \leq\epsilon_{0}|U\right\}
^{\mathrm{I}\left(  a_{0}=1\right)  }\Pr\left\{  h\left(  1,U\right)
\leq\epsilon_{1}|U\right\}  ^{\mathrm{I}\left(  \mathrm{a}_{1}=1\right)  }\\
&  \times\Pr\left\{  h\left(  0,U\right)  \leq\epsilon_{0}|U\right\}
^{\mathrm{I}\left(  a_{0}=1\right)  }\Pr\left\{  h\left(  1,U\right)
>\epsilon_{1}|U\right\}  ^{\mathrm{I}\left(  \mathrm{a}_{1}=0\right)  }\\
&  \times\Pr\left\{  h\left(  0,U\right)  >\epsilon_{0}|U\right\}
^{\mathrm{I}\left(  a_{0}=0\right)  }\Pr\left\{  h\left(  1,U\right)
\leq\epsilon_{1}|U\right\}  ^{\mathrm{I}\left(  \mathrm{a}_{1}=1\right)  }\\
&  \times\Pr\left\{  h\left(  0,U\right)  >\epsilon_{0}|U\right\}
^{\mathrm{I}\left(  a_{0}=0\right)  }\Pr\left\{  h\left(  1,U\right)
>\epsilon_{1}|U\right\}  ^{\mathrm{I}\left(  \mathrm{a}_{1}=0\right)  }\\
&  =\Pr\left\{  A^{z=0}=a_{0}|U\right\}  \Pr\left\{  A^{z=1}=a_{1}|U\right\}
\end{align*}
such that
\begin{align*}
&  \Pr\left\{  \mathcal{C=}\mathrm{co}|U,\mathcal{N}=1\right\} \\
&  =\frac{\Pr\left\{  A^{z=0}=0,A^{z=1}=1|U\right\}  }{\Pr\left\{
A^{z=0}=0,A^{z=1}=1|U\right\}  +\Pr\left\{  A^{z=0}=1,A^{z=1}=0|U\right\}  }\\
&  =\frac{\frac{\Pr\left\{  A^{z=0}=0,A^{z=1}=1|U\right\}  }{\Pr\left\{
A^{z=0}=1,A^{z=1}=0|U\right\}  }}{\frac{\Pr\left\{  A^{z=0}=0,A^{z=1}%
=1|U\right\}  }{\Pr\left\{  A^{z=0}=1,A^{z=1}=0|U\right\}  }+1}\\
&  =\frac{\frac{\Pr\left\{  A^{z=0}=0|U\right\}  \Pr\left\{  A^{z=1}%
=1|U\right\}  }{\Pr\left\{  A^{z=0}=1|U\right\}  \Pr\left\{  A^{z=1}%
=0|U\right\}  }}{\frac{\Pr\left\{  A^{z=0}=0|U\right\}  \Pr\left\{
A^{z=1}=1|U\right\}  }{\Pr\left\{  A^{z=0}=1|U\right\}  \Pr\left\{
A^{z=1}=0|U\right\}  }+1}\\
&  =\frac{\frac{\Pr\left\{  A^{z=1}=1|U\right\}  }{\Pr\left\{  A^{z=1}%
=0|U\right\}  }\frac{\Pr\left\{  A^{z=0}=0|U\right\}  }{\Pr\left\{
A^{z=0}=1|U\right\}  }}{\frac{\Pr\left\{  A^{z=1}=1|U\right\}  }{\Pr\left\{
A^{z=1}=0|U\right\}  }\frac{\Pr\left\{  A^{z=0}=0|U\right\}  }{\Pr\left\{
A^{z=0}=1|U\right\}  }+1}\\
&  =\frac{\exp\left\{  p\left(  1\right)  +U\right\}  \exp\left\{  -p\left(
0\right)  -U\right\}  }{\exp\left\{  p\left(  1\right)  +U\right\}
\exp\left\{  -p\left(  0\right)  -U\right\}  +1}\\
&  =\frac{\exp\left\{  p\left(  1\right)  -p\left(  0\right)  \right\}  }%
{\exp\left\{  p\left(  1\right)  -p\left(  0\right)  \right\}  +1}%
\end{align*}
does not depend on $U;$ therefore proving that $\mathcal{C}\amalg
U|\mathcal{N}=1.$

\end{document}